\documentclass[prl,aps,twocolumn,showpacs,amssymb]{revtex4}
\usepackage{color,graphicx,shortvrb}
\usepackage{amsfonts}
\usepackage{tabularx}
\usepackage{dcolumn}

\begin{document}
\title{Spontaneous Dissociation of $^{85}$Rb Feshbach Molecules} \author{S.~T. Thompson, E. Hodby, and C.~E. Wieman} \affiliation{JILA, National Institute of Standards
and Technology and the University of Colorado, and the Department
of Physics, University of Colorado, Boulder, Colorado 80309-0440}

\date{\today}

\begin{abstract}
The spontaneous dissociation of $^{85}$Rb dimers in the highest
lying vibrational level has been observed in the vicinity of the
Feshbach resonance which was used to produce them.  The molecular
lifetime shows a strong dependence on magnetic field, varying by
three orders of magnitude between 155.5~G and 162.2~G.  Our
measurements are in good agreement with theoretical predictions in
which molecular dissociation is driven by inelastic spin
relaxation.  Molecule lifetimes of tens of milliseconds can be
achieved close to resonance.
\end{abstract}


\maketitle Magnetic Feshbach resonances were first used to
dramatically alter the strength and sign of interatomic
interactions in ultracold
atoms\cite{Inouye,Wiemansetup,Jininteractions,Ketterleinteractions,Thomas,Salomoninteractions}.
Today they have become very useful tools for creating ultracold
gases of diatomic molecules.  In our initial experiments we saw
molecules formed from a $^{85}$Rb BEC by nonadiabatic mixing of
atomic and molecular states when the magnetic field was rapidly
pulsed close to the Feshbach resonance\cite{Wiemanoscillations}.
Subsequently it has been shown that both
fermionic\cite{Jinfirst,Hulet,Salomon} and
bosonic\cite{GrimmCs,KetterleNafirst,Rempe} atoms can be converted
into molecules by adiabatically sweeping the magnetic field
through a Feshbach resonance.  Molecules formed using these
techniques are very weakly bound and very highly vibrationally
excited and are of considerable experimental and
theoretical\cite{Burnett,Hutson,Shlyapnikov} interest.  Near the
Feshbach resonance the size of the molecules is comparable to the
interatomic spacing.  The lifetimes of these molecules has varied
widely under different conditions and their decay processes have
not been fully established.

Several experiments have shown that such molecules can undergo
rapid vibrational quenching in which they collide with atoms or
other molecules and relax to lower vibrational
states\cite{KetterleNasecond,Jinsecond}.  For the case of
molecules created from a Fermi gas, it has been observed that near
resonance the molecular lifetime increases by several orders of
magnitude\cite{Jinsecond}.  It is speculated that collisional
relaxation is greatly suppressed close to the Feshbach resonance
due to the Fermi statistics of the atoms\cite{Shlyapnikov}.  A
systematic study of the lifetime of molecules composed of bosons
near a Feshbach resonance has not yet been published.  However, it
is believed that the observed low atom-molecule conversion
efficiencies for bosonic atoms \cite{Rempe,KetterleNasecond} are
actually the result of very high vibrational quenching rates near
the Feshbach resonance.  In general, all of these experiments have
started with an atom cloud with an initial peak density of
10$^{13}$ -- 10$^{14}$ cm$^{-3}$.  This collisional quenching
mechanism will not be significant at lower densities, such as the
conditions we have used in studying the conversion of $^{85}$Rb
atoms to molecules.  However, in this Letter we show that the
molecular lifetime can be quite short even in the low density
regime.

Here we have systematically investigated the molecular lifetime of
$^{85}$Rb dimers in the highest vibrational state as a function of
magnetic field near the Feshbach resonance.  By starting with an
ultracold but uncondensed gas of bosonic $^{85}$Rb atoms in a
magnetic trap we have been able to study the molecular lifetime at
an initial atom density which is two to three orders of magnitude
smaller than in other experiments and thus distinguish collisional
destruction of molecules from the intrinsic lifetime of the
molecular state.  K\"{o}hler $et~al.$ have predicted that a very
different decay mechanism should dominate under these
conditions\cite{Julienne}.  In short, they expect inelastic spin
relaxation to lead to the spontaneous decay of these molecules.
One of the atoms in the molecule experiences a spin flip that is
similar to an inelastic spin relaxation collision between two
atoms.  This causes the molecule to dissociate, releasing
sufficient kinetic energy for both atoms to be lost from the trap.
A high dependence of this dissociation rate on magnetic field is
anticipated.  Close to resonance the size of the molecule
increases and spin relaxation is suppressed.  Our work directly
tests this theoretical prediction and determines the range of
experimentally accessible molecular lifetimes.

To carry out these lifetime measurements we start with what has
become a rather standard technique for molecule production, namely
ramping the magnetic field adiabatically through a Feshbach
resonance\cite{Jinfirst,Hulet,Salomon,GrimmCs,KetterleNafirst,Rempe}.
We have used the 11G wide resonance at 155G for this purpose and
have observed a 30\% atom-molecule conversion efficiency.  We
found the lifetime of the molecules by holding them for various
lengths of time, then converting all remaining molecules back into
atoms and measuring the number of atoms remaining versus the
duration of the hold.  We have repeated this process holding the
molecules at several different magnetic fields.

The apparatus used in this study has been described in detail
elsewhere\cite{Wiemansetup}.  We first prepared an ultracold (30
nK) thermal cloud of 100,000 $^{85}$Rb atoms in the $F = 2$, $m_F
= -2$ state in a magnetic trap at a bias field of 162.2~G.  The
standard deviation of the atom number from shot to shot was
$\sim$3\%.  The spatial distribution of the atoms was Gaussian
with a peak density of n$_0$ = 6.6~x~10$^{11}$~cm$^{-3}$ and the
trap frequencies were (17.5 x 17.2 x 6.8) Hz.   We then used the
trapping coils to apply a magnetic field time sequence as shown in
Fig.~1 to produce molecules and subsequently measure their
lifetime.

\begin{figure}
\includegraphics[bb=98 257 510 528,clip,scale=0.45]{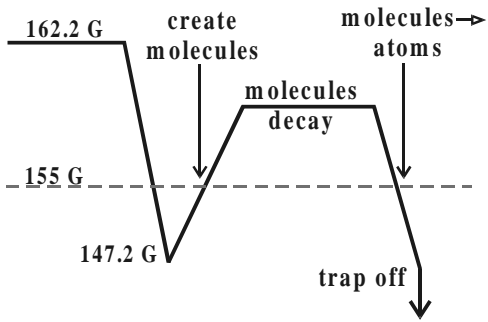}
\caption{Magnetic field ramp sequence for producing molecules and
measuring their decay rate.  The Feshbach resonance in indicated
by the dashed line.  The field is first swept as quickly as
possible from the evaporation field to the opposite side of the
resonance.  A second slower ramp back across the resonance
converts some atoms to molecules.  The molecules are then held at
a constant field above the resonance for a variable amount of
time.  A third ramp across the resonance then converts any
remaining molecules back into atoms and the magnetic trap is
turned off.}
\end{figure}

Having performed evaporative cooling at 162.2~G where the
scattering length is positive, we first ramped the magnetic field
to 147.2~G as rapidly as experimentally convenient (an inverse
ramp rate of 46 $\mu$s/G) simply to get to the correct side of the
resonance to begin molecule production\cite{Yurovsky}.  A second
slower ramp (57 $\mu$s/G) back across the resonance then
adiabatically converted 30\% of the atoms into molecules.  This
field ramp continued to the chosen field $B_{hold}$ above the
resonance.  The field was then held constant at $B_{hold}$ for a
variable amount of time $t_{hold}$, during which time a fraction
of the molecules could decay.  A third ramp across the resonance
(65 $\mu$s/G) then converted any remaining molecules back into
atoms.  The trap was then turned off and the atom cloud was
allowed to expand for 22~ms before destructive absorption imaging
was used to determine the number of atoms in the cloud.  By
measuring the decrease in the number of atoms as a function of
$t_{hold}$ we were effectively measuring the decay of the
molecules.  This method of course relies on the assumption that
the decaying molecules leave the magnetic trap so we don't see
them in our absorption images.  The observed exponential loss
indicates that this must be true for at least a large fraction of
them.  On theoretical grounds it is likely all leave since it has
been predicted that the decay energies associated with the various
available decay channels are all on the order of several
mK\cite{Julienne} and our trap depth is only $\sim$1~mK.  Also, we
have looked at absorption images at a large range of expansion
times and have not seen any evidence for modestly energetic atoms
arising from less energetic decay channels.  By measuring the atom
number as a function of $t_{hold}$ and by fitting this to an
exponential decay we were able to extract the molecular lifetime
at $B_{hold}$.

Data from such a measurement is shown in Fig.~2$a$ for $B_{hold}$
= 156.6~G.  We have investigated a range of $B_{hold}$ from 155~G
to 162.2~G.  The decay we observe fits very nicely to an
exponential.  We found that the time constant for the decay
depends very strongly on field; it changes by three orders of
magnitude over this 7~G wide region.

\begin{figure}
\includegraphics[bb=93 85 510 684,clip,scale=0.45]{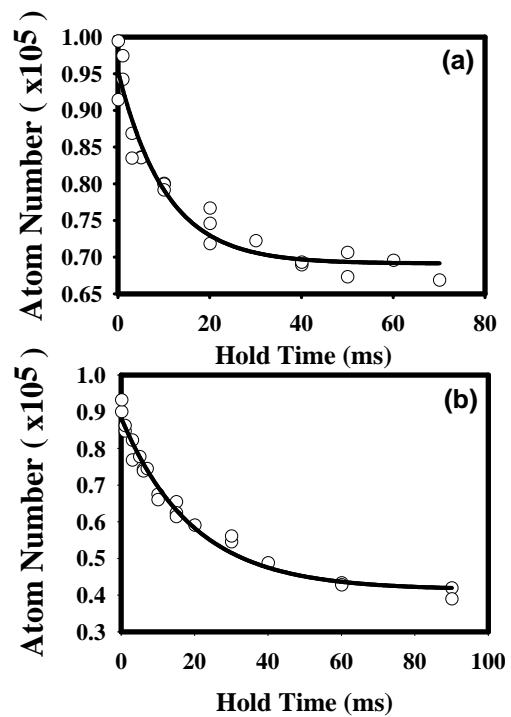}
\caption{Measurements of molecular lifetime.  (a) Number of atoms
remaining after holding the system at 156.6~G as a function of the
hold time.  The decay fits nicely to an exponential and from this
we get a lifetime of 10.4(1.7)~ms.  The baseline indicates we have
converted 30\% of the atoms into molecules. (b) Number of atoms
remaining after holding at 155.5~G as a function of the hold time.
In this case atoms are being lost from the trap during the hold
time due to three body collisions and we never observe a
horizontal baseline.  After correcting for this loss we get a
molecule lifetime of 24.7(6.4)~ms.}
\end{figure}

We have observed that for values of $B_{hold}$ within $\sim$1~G of
the Feshbach resonance the interpretation of the data is
complicated by the fact that some atoms also leave the trap during
$t_{hold}$ mostly due to three body collisions.  The exponential
decay in Fig.~2$b$ exhibits a decaying baseline due to this atom
loss.  To compensate for this atom loss we measured the loss of
atoms directly (no molecules present) at the appropriate densities
and magnetic fields and subtracted this loss from our raw
molecular decay data.  In this way the molecular lifetime close to
the Feshbach resonance was extracted.  A similar technique needed
to be employed by Regal $et~al.$ \cite{Jinsecond} in the
measurements of the lifetime of their $^{40}$K molecules produced
from a Fermi gas.  A summary of our molecule lifetime measurements
is shown in Fig.~3.  The three data points closest to the Feshbach
resonance show the atom loss correction described above.  For the
point at 156.6~G the correction is negligible. Since we know that
the three body loss rate decreases rapidly at higher
fields\cite{Wiemancollisions} we can safely ignore this atom loss
correction for all points above 156.6~G.

The solid curve in Fig.~3 is the result of a coupled channels
calculation done by K\"{o}hler $et~al.$ in ref.~\cite{Julienne} in
which inelastic spin relaxation leads to the spontaneous decay of
these molecules.  There is good agreement between experiment and
theory for fields greater than $\sim$157~G covering a factor of
100 in lifetime.  The discrepancy close to the Feshbach resonance
is most likely due to other decay processes not included in the
theory that may become significant close to resonance such as
atom-molecule or molecule-molecule collisions. The dashed curve is
the result a universal calculation which does not depend on the
detailed nature of interatomic interactions, also by K\"{o}hler
$et~al.$ in ref.~\cite{Julienne}.  It predicts that the molecular
lifetime as a function of magnetic field is given by
4$\pi$a$^3$(B)/K$_2$(B) where a(B) is the s-wave scattering length
and K$_2$(B) is loss rate constant for inelastic spin relaxation
collisions.  This simple formula also does a good job of
predicting the molecular lifetime over the magnetic field range we
have investigated and in addition provides good physical insight
into the decay mechanism.  It has been theoretically shown that
the spatial extent of the wave functions of these Feshbach
molecules is of the order of the scattering length\cite{Burnett}.
Thus, as a(B) becomes large near resonance so does the volume
containing the atom pair and the spontaneous decay of the molecule
is suppressed.  As pointed out in ref.\cite{Julienne}, if K$_2$(B)
is known, such measurements of the molecular lifetime can be used
as a direct probe of the size of the molecule.

\begin{figure}
\includegraphics[bb=75 215 544 548,clip,scale=0.5]{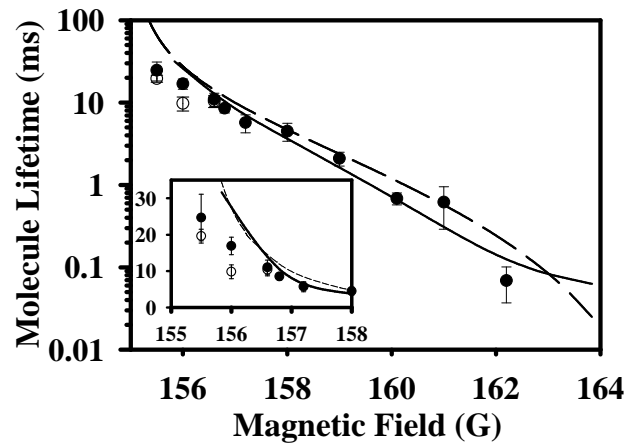}
\caption{Molecule lifetime as a function of magnetic field.  The
experimental data are represented by the closed points.  The open
points close to the Feshbach resonance are the raw data before the
atom loss correction was applied.  The two lines are the results
of theoretical calculations with $no~free~parameters$ by
K\"{o}hler $et~al.$ (ref.~19) in which molecules spontaneously
decay due to inelastic spin relaxation.  The solid line arises
from an exact coupled channels scattering calculation.  The dashed
line results from a simpler calculation in which the detailed
nature of the interatomic potentials is ignored, resulting in an
analytic solution for the molecular lifetime.  The inset shows the
discrepancy between experiment and theory close to the 155.04~G
Feshbach resonance.}
\end{figure}

In summary, we have measured the lifetime of $^{85}$Rb dimers in
the highest lying vibrational level in the vicinity of the
Feshbach resonance.  We have observed a very strong dependence of
this lifetime on magnetic field which is in good agreement with
theoretical predictions where molecules decay due to dissociation
driven by inelastic spin relaxation.  These results show that it
is possible to create $^{85}$Rb dimers with lifetimes of tens of
milliseconds.  These results also explain the unexplained
atom/molecule loss observed in our previous
experiments\cite{Wiemanoscillations,MHolland,BurnettoscillationsPRA,Burnettoscillationscondmat}
creating coherent superpositions of atomic and molecular BECs of
$^{85}$Rb.

We thank P. Julienne and T. K\"{o}hler for their theoretical
assistance and D. Jin and C. Regal for helpful discussions.  This
work has been supported by ONR and NSF.  S.T. Thompson
acknowledges support from ARO-MURI and E. Hodby acknowleges the
support of the Lindemann Foundation.

\end{document}